\documentclass[12pt,preprint]{aastex}

\lefthead{Cunha et al.}
\righthead{Chemical Evolution of the Bulge}

\begin{document}

\title{Chemical Evolution of the Galactic Bulge as Derived from High-Resolution 
Infrared Spectroscopy of K and M Red Giants} 

\author{Katia Cunha}
\affil{National Optical Astronomy Observatory, P.O. Box 26732, Tucson, AZ
85726, USA \\ Observat\'orio Nacional, Rua General Jos\'e Cristino 77, 
20921-400, S\~ao Crist\'ov\~ao, Rio de Janeiro, Brazil; kcunha@noao.edu}
\author{Verne V. Smith}
\affil{National Optical Astronomy Observatory, P.O. Box 26732, Tucson,
AZ 85726, USA; vsmith@noao.edu}

\begin{abstract}

We present chemical abundances in K and M red-giant members of the Galactic 
bulge derived from high-resolution infrared spectra
obtained with the Phoenix spectrograph on Gemini-South. 
The elements studied are carbon, nitrogen, oxygen, sodium, titanium,
and iron. The evolution of C and N abundances in the studied red-giants
show that their oxygen abundances represent
the original values with which the stars were born. Oxygen is a superior
element for probing the timescale of
bulge chemical enrichment via [O/Fe] versus
[Fe/H]. The [O/Fe]-[Fe/H] relation in the bulge does not follow
the disk relation, with [O/Fe] values falling above those of the disk.
Titanium also behaves similarly to oxygen with respect to iron.
Based on these elevated values of [O/Fe] and [Ti/Fe] extending
to large Fe abundances, it is suggested that the bulge underwent
a more rapid chemical enrichment than the halo. In addition, there are
declines in both [O/Fe] and [Ti/Fe] in those bulge targets with the largest Fe abundances, signifying
another source affecting chemical evolution: perhaps Supernovae of Type Ia. Sodium
abundances increase dramatically in the bulge with increasing metallicity, possibly reflecting the
metallicity dependant yields from supernovae of Type II, although Na contamination
from H-burning in intermediate mass stars cannot be ruled out. 

\end{abstract}



\section{Introduction}
The idea that galaxies consist of distinct stellar populations goes
back to the classic observations of M31, M32, and NGC 205 by
Baade (1944).  Within Baade's concept of populations, the Milky Way
can be divided somewhat crudely into the
halo, thick disk, thin disk, and bulge.  Understanding the ages,
kinematics, and chemical enrichment histories of the various
populations of a galaxy provides insight into and constraints on
models of galaxy formation and evolution.   The elemental abundance
distributions, in particular the abundance ratios of certain critical
elements in stars are sensitive to global variables such as star
formation histories and chemical enrichment  timescales within
the various galactic components.

Within the Milky Way, there are numerous abundance studies of the
disk (e.g., Edvardsson et al. 1993; Reddy et al. 2003), thick disk
(e.g. Prochaska et al. 2000; Bensby et al. 2003), or halo (e.g., Fulbright 2002; Fulbright 
\& Johnson 2003),
all of which provide an increasingly detailed view of chemical
evolution and evolutionary timescales in the Galaxy.
Noticeably absent from the many lists of detailed abundance studies are
corresponding abundance analyses of the bulge population.
This is due to its distance ($\sim$8 kpc) and interstellar absorption,
both of which combine to render bulge stars, even the bright K and M
giants, relatively faint for high-resolution spectroscopic analyses. 

The earliest determinations of abundances in bulge stars were
derived from low-resolution spectra and consisted of overall metallicities
as characterized by [Fe/H].  These initial abundance studies were
by Rich \& Whitford (1983), Rich (1988; 1990), Terndrup, Sadler \&
Rich (1995), or Sadler, Rich, \& Terndrup (1996).
The early heroic study by McWilliam \& Rich (1994) was the pioneering
effort to probe elemental abundance distributions and chemical
evolution in the bulge population, but suffered from somewhat low
spectral-resolution and signal-to-noise (S/N).

The dearth of bulge high-spectral resolution abundance studies is
now being remedied by prototype studies that
have much higher-quality spectra at their disposal.  The most
recent works are the
detailed study of stellar parameters and iron abundances in
27 bulge K-giants by Fulbright, McWilliam \& Rich (2006), or the
analysis of 14 M-giants by Rich \& Origlia (2005).  Fulbright et al.
(2006), in their first paper of a planned series, focus on
carefully deriving
the fundamental stellar parameters of effective temperature, surface
gravity, iron abundance, and microtubulent velocity.  Their discussion
then centers on comparing the iron abundances in their 27-star
sample with the same stars in the larger, low-spectral resolution
studies of Rich (1988) and Sadler et al. (1996).   By transforming
the Fe-abundances from the older (but with larger stellar samples)
studies onto their Fe-abundance scale as set by their sample
from high-quality, high-resolution spectra, Fulbright et al. derive
an updated metallicity (i.e. [Fe/H]) distribution for the bulge.
Rich \& Origlia (2005) concentrate on bulge M-giants that have
a rather restricted range in metallicity ([Fe/H] $\sim$ -0.3 to 0.0)
but the sample size in this range allows them to evaluate the scatter 
in abundance ratios. 

In this study, we present abundances 
in 7 bulge red-giant members; 5 K-giants and 2 M-giants. Our spectral
types thus overlap with Fulbright et al. (2006) and Rich \& Origlia (2005),
as well as spanning a wide metallicity range. We are thus in a
position to probe the chemical evolution of a small, but key set
of elements in the bulge. 

\section{Observations}

The target stars in this bulge study are a subset from the  
K-giants sample analyzed by Fulbright et al. (2006), plus two additional
M-giants (one being in common with Rich \& Origlia 2005).
High-resolution infrared spectra were obtained with the Gemini South
Telescope and the Phoenix spectrograph (Hinkle et al. 2003) during
queue observing in May 2004, July 2004, June 2005, and July 2005.
The observations were obtained at different positions on the slit and
followed the same scheme described in Smith et al. (2002). 

The spectra obtained are single order echelle with a 
resolution R=$\lambda$/$\Delta$$\lambda$=50,000 (corresponding 
to a resolution element of $\sim$ 4-pixels). The observations
of the K-giants were obtained at two grating tilts,
one in the K-band region ($\lambda$ $\sim$ 23,350 \AA), and 
another in the H-band region ($\lambda$ $\sim$ 15,540 \AA).
Due to meteorological conditions, the M-giant targets were only observed in the H-band region.
For those observations obtained
in the K-band, spectra of hot stars were also needed for telluric division.
The observed spectra were reduced to one-dimension by means of 
standard sets of IRAF routines (see Smith et al. 2002 for details). 
The typical S/N of the observed spectra is 100-200, with one star (IV-167) having
a S/N = 50.
A reduced Phoenix spectrum in the K-band region is shown in Figure 1 for target star
IV-003; the prominent features appearing in the spectra, such as the
numerous C$^{12}$ O$^{16}$ lines and the Na I line are identified.

\section{Analysis}

The stellar parameters of effective temperature (T$_{\rm eff}$) and surface
gravity (parametrized as log g) are required for elemental abundance
determinations. In addition, an estimate of the microturbulent velocity 
parameter, $\xi$, and the overall metallicity of the model atmospheres adopted
in the calculations are needed.

\subsection{Reddening, Photometric Colors, and Effective Temperatures}

The spheroidal component of the Milky Way Galaxy has not been studied extensively 
due to the extremely high visual extinction ($\sim$ 30 mag) 
that affects its bulge.
Even for target stars lying in Baade's window, where there is significantly less
extinction, it is challenging to obtain intrinsic (de-reddened) colors to an
accuracy of $\sim$0.01 mag, which is needed in order to derive effective
temperatures accurate to within 50K.
Stanek (1996) constructed extinction and reddening maps for  
Baade's window from color-magnitude diagrams
obtained by the Optical Gravitational Lensing Experiment (OGLE; Udalski et al. 1993, 1994). 
These maps showed that extinction in Baade's window is quite irregular, varying
between 1.3 and 2.8 mag in A$_{v}$, with an estimated error of $\sim$ 0.1 mag.
We used the extinction maps of Stanek (1996) and the A$_{v}$'s which
are listed in Table 1 to derive the de-reddened colors for the targets stars (also in Table 1). 
The V magnitudes were taken from  
OGLE and the infrared magnitudes were taken from the 2MASS database.   

In order to obtain effective temperatures for the K-giants, 
we adopted the Alonso et al. (1999) T$_{eff}$-calibrations
for (J-K) and (V-K) colors, which are defined in the TCS (Telescopio Carlos Sanchez) system. 
The 2MASS colors for the target stars were transformed to the TCS system
using the relations presented in Ramirez \& Melendez (2005). 
Although the (V-K) calibration is considered one of the best T$_{eff}$ indicators for cool stars 
(with an internal error estimated by Alonso et al. to be of $\sim$ 25K), 
the (V-K) color is also more sensitive to reddening, which represents a potential problem in
bulge studies. On the other hand, the (J-K) color is less sensitive to reddening
but the (J-K) T$_{eff}$-calibration has a larger internal error of $\sim$125 K 
(according to Alonso et al. 1999). Given these uncertainties, 
we adopted an average between the T$_{eff}$'s obtained from (J-K) and (V-K) calibrations
as the final temperatures for the sample K giants. The average 
T(V-K) - T(J-K) for our sample is -48 $\pm$ 90K and we note that there is no trend with A$_{v}$.

The T$_{eff}$-calibrations adopted for the K-giants are not defined for the cooler
giants of M spectral types. For the two M-giants in our sample we adopted the (J-K) calibration from 
Houdashelt et al. (2000). Transformations of the 2MASS colors to the CTIO
system were needed, and we used the relations in Carpenter (2001).
The Houdashelt et al. (2000) calibration has a small 
metallicity and log g dependance: for log g varying between 0.5 and 1.0 dex,
(J-K) varies less than 0.02 mag. The calibration for log g=1.0 and [Fe/H]=0.0
are roughly coincident with the one for logg=0.5 and [Fe/H]=0.25.
We adopted an average calibration for  
the parameter space between log g=0.5 and 1.0 (giant stars) and metallicities
around solar (given that the target M giants have TiO bands and that very metal poor
M-giants would not exist).
Derived stellar parameters for target stars are presented in Table 1. As a consistency
check we also estimated the effective temperature for the hotter M giant, BMB78, by means
of the Bessel et al. (1998) calibration: this yields T$_{eff}$=3570 K, which is in
good agreement with the value listed in Table 1.

\subsection{Surface Gravities and Microturbulent Velocities}

Classical determinations of spectroscopic log g's rely on the forced agreement
between the abundances obtained from Fe I and Fe II lines, typically, in the optical.
Our observed infrared spectra do not contain adequate
Fe lines in order to validate a purely spectroscopic determination.
Surface gravities, however, can also be estimated from standard relations between stellar 
luminosity and mass. The log g's in this study were estimated using the derived 
T$_{eff}$'s in Table 1 and the corresponding positions of the stars relative to isochrones computed
by Girardi (2000), iterated for the final values of metallicities for
each star. Since the bulge of the Milky Way
is an old population we adopted the isochrones corresponding to 10 Gyr, and we note that
age differences of several Gyrs will not have significant effects on the log g values
for these old red-giants. 

Microturbulent velocities ($\xi$) for giants can be estimated from the set of 
molecular CO lines which are present in the observed K-band spectra and that
span a range in line strengths. Equivalent widths were measured for several CO lines (Table 2)
and microturbulences for the K-giants were obtained from the requirement that the abundance is
independant of the measured equivalent width. For the M-giants the microturbulences
were estimated from the overall match between observed and synthetic spectra
at 1.55 $\mu$m.  

The adopted microturbulences (listed in Table 1)
vary between $\xi$= 1.8 and 2.2 km/s for the K-giants; for the two M-giants in our sample we obtained 
$\xi$ = 2.5 and 3.0 km/s. These values are consistent with the trend of 
microturbulence
versus the bolometric magnitudes (M$_{bol}$) and microturbulence versus log g's
found for Galactic M giants (Smith \& Lambert 1985, 1986, 1990) and LMC red-giants 
(Smith et al. 2002). As discussed 
in Smith et al. (2005), the microturbulent velocities derived from infrared CO lines 
seem to be systematically higher than the ones typically obtained from Fe I lines in the optical.
We note that the microturbulences derived here for the K-giant targets are also measurably 
larger than the values
obtained from the Fe I lines in the optical by Fulbright et al. (2006) for the same stars;
the average difference is 0.6 km s$^{-1}$.
A comparison with the study by Rich \& Origlia (2005) is not possible because the latter adopted
a constant microturbulence of 2 km/s in their abundance analyses of M-giants.

\subsection{Abundances}

In the two wavelength intervals covered by the Phoenix spectra there are a number of
CO, OH, and CN transitions, but only a few atomic absorption lines. 
We derived abundances of carbon,
nitrogen and oxygen from molecular transitions of CO, CN and OH.
In addition, we analyzed atomic lines from Na I, Ti I and Fe I.
In order to derive the abundances we calculated synthetic spectra
using the LTE synthesis code MOOG (Sneden 1973). All abundances were derived via spectrum
synthesis, however, Table 3 displays the lines of OH, CN, Fe I, Na I and Ti I that
were used for the abundance determinations.
The molecular and atomic transitions analyzed in each case are discussed in the
following section.

The model atmospheres used in the abundance calculations were interpolated from the 
original grid of MARCS models (BEGN, Gustafsson et al. 1975).
The models were computed intially for metallicities ([Fe/H]) taken from 
Fulbright et al. (2006) and Rich \& Origlia (2005). These metallicities
were then adjusted until the metallicities of the models
and the derived Fe abundances (from our sample Fe I infrared lines) were consistent.
(The derivation of the Fe abundances from our infrared spectra will be discussed in
the next section). A comparison of
the model atmospheres adopted in this study, MARCS, with the more modern
SOSMARCS (Plez, Brett \& Nordlund 1992) indicates that differences in 
abundances are not significant (see Cunha et al. 2002).

\subsubsection{Iron}

Iron deserves discussion first as it is the fiducial element that is usually
used as the primary 'metallicity' indicator. The study here utilizes
only IR spectra with limited wavelength coverage and iron is represented
by a few well-defined Fe I lines near $\lambda$ 15530-15540 $\AA$ (Table 3). 
These Fe I lines are readily measurable in the Sun plus the well-studied
K-giant $\alpha$ Boo. No laboratory measurements of oscillator strength
exist for these lines but semi-empirical values are available from the Kurucz
line list (Kurucz \& Bell 1995), as well as solar gf-values from Melendez \& Barbuy
(1999).

In order to set the gf-values for Fe I lines, we measured equivalent widths
from the disk-integrated solar spectrum from (Wallace et al. 1996) and used both Kurucz and 
Melendez \& Barbuy (1999) gf-values to determine the solar photospheric iron
abundance. Also, two solar models were used: one from the MARCS code and
one from Kurucz ATLAS9 with T$_{eff}$=5777K, log g=4.438, and microturbulent
velocity, $\xi$= 0.8 km/s. Note that the most recent photospheric abundance
as reviewed by Asplund, Grevesse \& Sauval (2005) is A(Fe)=7.45. 
The Fe I lines from Table 3 yield A(Fe)=7.31 $\pm$ 0.18 for the MARCS
model and 7.37 $\pm$ 0.13 with the ATLAS9 model when Kurucz gf-values
are used. We prefer to anchor our
Fe analysis to a solar abundance of A(Fe)=7.45 with the MARCS model and
adjusted the individual gf-values so that each line yielded 7.45 for A(Fe).

Given the solar gf-values defined above, an analysis of $\alpha$ Boo, using
the spectra from Hinkle et al. (1996), provides
a test of what Fe abundance these Fe I lines plus gf-values would yield 
for this well-studied K-giant. Examples of previous uses of  $\alpha$ Boo
in this role can be found, for example, in Smith et al. (2000) or Fulbright et al.
(2006). Smith et al. derived T$_{eff}$=4300K log g=1.7, $\xi$= 1.6 km/s, and [M/H]=-0.60, while
Fulbright et al. (2006) find T$_{eff}$=4290K, log g=1.55,  $\xi$= 1.67 km/s, [M/H]=-0.50.
The $\alpha$ Boo analysis used the same Fe I lines (and solar gf-values) that were
adopted in this study of bulge red-giants, along with the model atmosphere as 
defined by Smith et al. (2000).
The abundance results for $\alpha$ Boo are listed in Table 4:
we find A(Fe) = 6.96 $\pm$ 0.05 ([Fe/H]=-0.49 $\pm$ 0.05). This value of [Fe/H]
for $\alpha$ Boo is in excellent agreement with Fulbright et al. (2006) of
[Fe/H]$_{\alpha Boo}$=-0.49 $\pm$ 0.07, and suggests that our bulge Fe abundances
scales can be compared directly at the low metallicity end. 

\subsubsection{Sodium and Titanium}

In addition
to Fe, as discussed in the previous section, there is a well-defined Ti I line
at $\lambda$ 15543.720$\AA$ and one for Na I at $\lambda$ 233379.137\AA. Both
of these atomic lines are measurable in the Sun and we have (as with Fe I lines)
derived solar gf-values using the 1D MARCS solar model with $\xi$=0.8 km s$^{-1}$ and
A(Na)=6.17 and A(Ti)=4.90 (Asplund et al. 2005). These are the gf-values listed
in Table 3.

With our derived solar gf-values the analysis of $\alpha$ Boo yields A(Na)=5.86 and
A(Ti)=4.83 (Table 4), with both abundances in agreement with abundances from optical
spectral lines. For example, Smith et al. (2000) find A(Na)=5.85$\pm$0.13 and 
A(Ti)=4.64 $\pm$ 0.12 based upon samples of optical lines of Ti I and Ti II. The agreement between
optical and IR is excellent for Na and within the errors for Ti.

\subsubsection{Carbon, Nitrogen and Oxygen}

The observed spectra around $\lambda$23400$\AA$ contain a number of first overtone vibration 
rotation
molecular CO lines. The selected CO lines are listed in Table 2, with the
excitation potentials ($\chi$) and gf-values for the transitions taken
from Goorvitch (1994). As discussed in Section 3.2, the sample CO lines 
were used to define the microturbulence parameter. We opted for measuring the
equivalent widths of the CO lines (Table 2) and obtained estimates for the microturbulent
velocities
from the independance of CO abundances versus measured equivalent widths. 
We then computed synthetic spectra of the regions containing the sample CO lines for
that derived microturbulent velocity. Fits to the observed profiles were obtained only
from adjustments of the CO abundances; as expected the selected microturbulences
were able to produce an overall good fit for strong and weak CO lines indicating 
that the adopted microturbulences and CO abundances were quite consistent with 
the results from synthetic fits of the entire CO spectral region.

The molecules of CO, OH, as well as CN, are abundant
and important constituents of the molecular equilibrium in the atmospheres of cool stars;
a consistent solution for the elemental abundances of carbon, nitrogen and oxygen has
to be iterated until a good match between synthesies and observations is achieved for
all of those spectral regions containing molecular transitions of CO, OH and CN, simultaneously.
In order to obtain carbon and oxygen abundances we also synthesized, at the same time, the 
spectral region around 15540 \AA, where there are 4 unblended lines of molecular OH.
The atomic data for the OH transitions needed in order to construct a linelist for the synthesis
were taken from Goldman et al. (1998) and the dissociation energy D$_{0}$ = 4.392 eV
from Huber \& Herzberg (1979) was adopted. A simultaneous solution was then obtained from the
best fit to the observed spectra containing CO and OH, that defined the elemental carbon and oxygen
abundances. 
Nitrogen abundances could then be obtained from weaker molecular CN lines molecular,
which are present around 15540 \AA, given that the carbon abundances are known.
The line list adopted for the calculation of the synthetic spectra for the CN region was compiled
B. Plez (private communication). The dissociation energy adopted is of 7.77 eV (Costes
et al. 1990). 

\subsubsection{Final Abundances and Their Uncertainties}

The atomic and molecular lines described in the previous three subsections were
used, in conjunction with the stellar parameters, in Table 1 to derive the
abundances listed in Table 4. Final abundances are presented for
$^{12}$C, $^{14}$N, $^{16}$O, Na, Ti and Fe, all based upon spectrum synthesis.
An example of spectrum sysnthesis of shown in Figure 2, which illustrates 
the Na I line in two bulge giants.

The S/N of the observed spectra is typically 100-200, with one star (IV-167) having
a S/N = 50. From Cayrel's (1988) formula, we estimate that the typical equivalent width 
uncertainty to range from $\sim$ 2 m\AA (e.g. for the CO line at 23367 \AA 
in I-322 we measure an equivalent width of 384 m\AA and the spectrum has a S/N=200) 
to 10 m\AA (for IV-167 with the lowest S/N in the sample) with a corresponding
abundance uncertainty between 0.01 and 0.04 dex. These errors are much smaller
than the abundance uncertainties caused by the errors in the stellar parameters
themselves.

An analysis of the uncertainties in the abundances due to errors in the fundamental
stellar parameters was conducted for each species. This analysis consisted
of perturbing a representative star and analyzing it with different model
atmospheres modified in Teff, log g, and $\xi$. In this case, the star IV-072
was used as the test and separate analyses were done for changes of
$\delta$T=+75K, $\delta$log g=$\pm$0.25 and $\delta \xi$ = $\pm$ 0.3 km s$^{-1}$.
The changes in the derived abundances for each element, as a function of each
fundamental parameter are listed in Table 5. The final estimated error for each
element is the quadrature sum of each uncertainty due to $\delta$T, $\delta$log g,
and $\delta \xi$; this error is labeled $\Delta$ and is shown in the last column
of Table 5. These errors dominate line-to-line scatter, in those cases where
more than one of two lines are available (such as CO), thus uncertainties
in the fundamental stellar parameters dominate the expected errors.  

In addition to changing the stellar parameters independently, a simultaneous
change in T$_{\rm eff}$ and log g is also investigated.  This is because
effective temperature and gravity are correlated along an isochrone, with
log g decreasing if T$_{\rm eff}$ is decreased, for example.  The last column
of Table 5 quantifies this effect.  Along these isochrones for old populations, a change of
75K in T$_{\rm eff}$ results in a change of about 0.15 dex in log g and it is
these changes that are shown in the last column of Table 5.  Again, IV-072 is
used as the test and its basic model of T$_{\rm eff}$=4400K and log g=2.4 is
changed to T$_{\rm eff}$=4325K and log g=2.25.  This change results in a small
change in $\xi$ as derived from the CO equivalent widths (from 2.2 to 2.1
km-s$^{-1}$) and the tabulated changes in the various elemental abundances.  These
changes are typical of what is found from the first exercise of changing the
paramerters independently and the errors listed in Table 5 are approximately
what is expected from an abundance analysis of this type.

As discussed in Section 3.3.1 our Fe abundance scale is close to the scale
of Fulbright et al. (2006). This comparison is illustrated in Figure 3
with Fe abundances plotted for the 5 bulge K-giants in common. The
overall agreement is good, although there may be a slight metallicity
dependant trend. The average difference and standard deviation 
$\Delta$A(Fe)(Us - Fulbright) = +0.11 $\pm$ 0.20. It is worth noting
that our estimated errors in stellar parameters will cause an uncertainty
of $\pm$0.12 dex in our Fe I abundances (Table 5). If this is combined with similar
errors from Fulbright et al. (2006), the combined scatter will be
$\pm$0.17 dex.
One of the M-giants, BMB 289, was also in the Rich \& Origlia (2005)
study and four elemental abundances are in common: $^{12}$C, O, Ti and Fe. 
The average difference (Us - Rich \& Origlia) and standard deviation of
the two analyses of BMB 289 is $\Delta$=+0.08 $\pm$ 0.17. These differences
can be considered to be within the expected uncertainties in the analyses.


\section{Discussion}

\subsection{Carbon, Nitrogen and Mixing in the Bulge Red Giants}

During stellar evolution along the red giant branch the abundance of carbon
and nitrogen change due to the extensive convective envelope of the giant
dredging material exposed to the CN-cycle to the spectroscopically 
observable surface. The main nuclei involved in the CN-cycle transmutations
are $^{12}$C, $^{14}$N, and $^{13}$C (with $^{15}$N being a very minor 
constituent). This phase of stellar evolution has been labelled 'first dredge-up',
and extensively studied both observationally and theoretically. A simple
sketch of first dredge-up would highlight the expectation
that $^{12}$C nuclei are converted to mostly $^{14}$N, with some
gain in the $^{13}$C abundance, and the $^{12}$C/$^{13}$C ratio
declines. If only the CN-cycle has operated significantly in the
stellar matter the total number of C+N nuclei will be conserved,
thus, as $^{12}$C is converted to $^{14}$N the $^{12}$C/$^{14}$N
will be lowered along a nearly predictable relation.

The target red giants are all evolved enough that they should have
undergone some amount of first dredge-up. The combination of $^{12}$C
and $^{14}$N abundances derived here can be used to quatify the
consequence of first dredge-up. 
Conservation of C+N nuclei allows a check on the metallicity, i.e.
Fe abundance, dependence of [C+N/Fe]. Ignoring $^{13}$C,
a minor constituent, as we have no $^{13}$C information from
the observed spectra, calculation of [$^{12}$C+$^{14}$N/Fe] reveals
no upward or downward trend: $^{12}$C+$^{14}$N scales as
solar with Fe. The average [$^{12}$C+$^{14}$N/Fe]=0.00 $\pm$ 0.18 with
the scatter of 0.18 dex being consistent within the uncertainties
of combining three individual elemental abundances into one ratio.

Evidence of the first dredge-up in the bulge red-giants is shown
in Figure 4; the top panel plots A($^{14}$N) versus A($^{12}$C).
The wavelength intervals observed in this study did not include
measurable features with  $^{13}$C, so the  $^{12}$C/$^{13}$C
is not known. The solar point is plotted here with the solid straight
line delineating [$^{12}$C/Fe] and [$^{14}$N/Fe] equal to 0.0. All of
the studied bulge red giants fall above the scaled solar line, i.e.,
to the $^{14}$N-rich zone. The solid curves show lines of constant
$^{12}$C+$^{14}$N, or the so-called 'CN-mixing lines'. Although the
$^{13}$C is not known, its maximum abundance due to H-burning
will be only $\sim$1/3.5 that of $^{12}$C, which is itself depleted
in mixed giants. The result of accounting for $^{13}$C in Figure 4
would flatten slightly the mixing curves as some $^{12}$C goes
into $^{13}$C instead of $^{14}$N. Nevertheless, the mixing curves
provide good approximations to what is observed in the bulge giants.
Each of the three curves at the lower $^{12}$C abundances (for
IV-003, IV329, and I-322) begin with initial $^{12}$C and $^{14}$N
scaled by the respective Fe abundances. The most $^{12}$C- and
$^{14}$N-rich mixing line begins an average Fe-abundance
representative of the most metal-rich stars. The position of the bulge red giants is
similar to what has been found for Galactic disk K and M giants
(Smith \& Lambert 1990) or even field M-giants in the LMC
(see Figure 8 in Smith et al. 2002). 

The $^{12}$C-depletion and $^{14}$N-enhancements
are indicative of CN-cycle material. The conclusion of this investigation
of the $^{12}$C and $^{14}$N abundances is that the $^{16}$O abundances
in these bulge red giants are not measurably altered from their primordial
values, as the mixing is not nearly extensive enough to have altered
the initial $^{16}$O measurably.

The bottom panel of Figure 4 shows another view of CN-cycle mixing with all
of the bulge giants scaled to a single mixing line. Here a solar $^{12}$C/$^{14}$N=3.16
is taken as the initial value and this ratio declines as $^{12}$C is converted
to $^{14}$N (or, as [$^{12}$C/Fe] declines). The mixing curve is a reasonable
approximation to the observed giants. Alpha Boo is also plotted with its
$^{12}$C, $^{14}$N, and Fe abundances derived from the same lines as used for
the bulge giants. This comparison bolsters confidence in the analysis and
the interpretation that this sample of bulge stars are first dredge-up 
red giants.

\subsection{The Oxygen Abundances and the Behavior of [O/Fe]}

In red giants the C and N abundances are excellent monitors of dredge-up
on the red giant branch.  In old, low-mass red giants, such as the ones
studied here, only the surface abundances of the
isotopes of C and N, along with the minor oxygen
isotopes $^{17}$O and $^{18}$O will be altered measurably by internal
mixing.  When studying low-mass giants it must be noted, however,
that significant fractions of globular cluster giants do show $^{16}$O
depletions that result from destruction by the higher temperature ON
parts of the CNO cycles.  The oxygen abundance variations are also
observed in globular cluster turn-off and main-sequence stars (e.g.
Gratton et al. 2001), thus the globular cluster O-variations arise from
chemical evolution seemingly peculiar to them (Smith et al. 2005;
Ventura \& D'Antona 2005).
The agreement in the bulge giants with the simple CN-mixing picture,
as discussed in the previous section, demonstrates that the $^{16}$O
abundances have not been altered measurably due to red giant
dredge-up.

With confidence that the bulge red giant oxygen abundances represent
the original values with which the star was born, they provide one
of the key monitors of bulge chemical evolution.  Oxygen has basically
a single nucleosynthetic origin from massive stars and is produced
over short timescales.  As the iron yields from SN II are much less than
oxygen, with Fe being synthesized most effectively in SN Ia over a
significantly longer timescale, the ratio of O/Fe is the best indicator of
SN II to SN Ia activity.

The Galactic disk and halo populations have had numerous published
studies with [O/Fe] versus [Fe/H] and its overall behavior is fairly
well-defined.  This is illustrated in the top panel of Figure 5, with
[O/Fe] taken from a number of studies (Edvardsson et al. 1993; Cunha, Smith \& Lambert
1998; Melendez, Barbuy \& Spite 2001; Smith, Cunha \& King 2001; 
Melendez \& Barbuy 2002; Fulbright \& Johnson 2003).                                        
The asterisks in this figure are specific stars identified as thick disk
stars from  Bensby, Feltzing, \& Lundstrom (2004).  The decreasing
values of [O/Fe] as [Fe/H] increase, for [Fe/H]$>$-1, is due to the
increased Fe production from the onset of SN Ia.  Below [Fe/H] of
$\sim$ -1.0 all Galactic stars exhibit an elevated value of [O/Fe].  The
simple exercise of combining oxygen and iron yields (with the caveat
that the Fe yields are uncertain in the models) from the grid of models
by Woosley \& Weaver (1995), convolved with a standard IMF results
in a value of [O/Fe]$\sim$+0.5: not very different from the observed
halo values.

The beginning of the  decrease in [O/Fe] in the Galactic halo occurs at
roughly [Fe/H]=-1.  The overall metallicity a population reaches (caused 
by SN II) when SN Ia's begin to contribute to the chemical evolution
is sensitive to the efficiency at which gas has been turned into stars.
In this simple picture, the offset between the thin and thick disks in
the [O/Fe]--[Fe/H] plane is caused by early, higher star formation
rate in the thick disk, relative to the thin disk, with the result that the
thick disk was more metal-rich by the time SN Ia's began to
explode.  The solid curve in the top panel of Figure 5 is a simple
numerical model of chemical evolution assuming no outflows or
infalls and instantaneous recycling.  The model has been adjusted 
to approximately follow the evolution from halo to thin disk.  The
thick disk stars from Bensby et al. (2004) are offset from this curve,
with the dashed line a linear fit to the thick disk points.

Bulge values of [O/Fe] versus [Fe/H] are plotted in the bottom panel 
of Figure 5 and can be compared to the halo and disk populations.
Red giants from this study are the large filled pentagon while the open squares
are the recent Rich \& Origlia (2005) M-giant abundances.  The advantage
of this study is the wider range in metallicity of targets, while the
strong point of Rich \& Origlia is the fairly large number of stars in a
limited metallicity range (providing some limits on the intrinsic
scatter in [O/Fe] in the bulge). At [Fe/H]$\sim$ -0.2 to 0.0, the
[O/Fe] trend indicated by the abundances obtained here seems to be slightly lower 
(on average by 0.18 dex) than
the results by Rich \& Origlia (2005);  within the uncertainties, however,
the abundances in the two studies roughly overlap. The halo and thin disk model
curve and the thick disk line are also plotted as references.  The
lowest-metallicity giant, IV-003 at [Fe/H]=-1.1, has a [O/Fe] value
that is consistent with the Galactic halo.  Among the more metal-rich
bulge giants, there is a significant decline in [O/Fe] relative to star
IV-003; the bulge population has undergone evolution in its
O/Fe ratios.  The decline in [O/Fe] in the bulge is not as large as for
the thin disk, but is reminiscent of the thick disk trend.  The most
straightforward explanation of this trend is that the bulge underwent
more rapid metal enrichment than the halo, but that star formation
did continue over timescales that may include the onset of SN Ia.

\subsection{Sodium in the Bulge}

The chemical evolution of oxygen, the subject of the previous section,
is dominated by SN II and its production is dependent on the mass
function of its high-mass parent stars.  One strong point of using
oxygen to track the formation history of massive stars, with the
subsequent dispersal of heavy elements and enrichment of the ISM, is
that O-yields from SN II are not very sensitive to the progenitor-star
metallicities.  The integrated yields of oxygen are determined mostly
by the shape of the stellar mass function: the ratios of very massive
to less massive SN II.

Moving up the Periodic Table to study sodium provides different, and
in some ways complementary insights into the details of chemical
evolution.  In most stellar populations studied to date Na is, like
O, primarily a product of SN II (Clayton 2003).  In globular clusters this may
not be the case, but we will discuss the globulars later in this
section.  The key to sodium's complementarity to an $\alpha$-element
such as oxygen is that the yields of Na from SN II are quite sensitive
to the inital SN II metallicities.  The main source of Na is
carbon-burning, but at the same time it can be destroyed at these
temperatures by proton captures.  The final yield from SN II thus
depends on the p/n ratio, which is itself a function of metallicity,
decreasing as the overall metallicity increases.

Figure 6 illustrates the behavior of Na and O in various stellar
populations and betrays their origins by plotting [Na/O] versus [O/H].
The top panel shows results for field stars of the Milky Way halo,
thick disk, and thin disk as small open circles, with these abundances
from Reddy et al. (2003); Fulbright (2000); Fulbright \& Johnson (2003);
Nissen \& Schuster (1997); Bensby, Feltzing \& Lundstrom (2003); Prochaska et al. (2000). 
The small filled symbols show [Na/O]
and [O/H] for the dwarf galaxies Sculptor (circles; Shetrone et al. 2003;
Geisler et al. 2005), Carina (triangles; Shetrone et al. 2003),
Fornax and Leo (pentagons; Shetrone et al. 2003), and the LMC (squares;
Smith et al. 2002).  There are two key points to
the top panel: 1) all of the various populations overlap within the
estimated errors of the many analyses (with the errors shown as
representative errorbars), and 2) these diverse populations define
a sequence of increasing Na/O with increasing metallicity (O/H).
The scatter in [Na/O] at a given metallicity is not much larger, if
at all, than the expected abundance uncertainties.  The nearly unique
relation of Na/O with O/H is dominated by SN II nucleosynthesis
driving their evolution.

The large filled pentagons in the top panel of Figure 6 are the
abundance results for the bulge stars.  This sample includes
a few quite metal-rich giants and provides values of [Na/O] at
large [O/H], plus sample stars overlap with the less O-rich populations.
In the overlap region of [O/H] the bulge stars agree well in
[Na/O] versus [O/H] with the other populations.  
Near solar metallicity ([O/H] $\sim$ 0.0), the bulge stars
overlap with the field-star sample but the lowest metallicity
bulge star in our sample (IV-003) falls well below
the trend. The Na I line in this star is very weak and the low
Na abundance does not result from incorrect stellar parameters.
(See the Na I synthesis in Figure 2.)
Since the Na/O yields from SN II are metallicity sensitive,
the low value of Na/O for IV-003 may indicate that the initial
enrichment of the bulge to [O/H] $\sim$ -1 (or, [Fe/H] $\sim$ -1.3)
was very rapid and dominated by very metal-poor massive stars
(with [m/H] less than -2). This would be an example in which
a metallicity-dependant ratio (Na/O) retained the metallicity
signature of its parent SN II, regardless of how polluted the
primordial cloud had become from the SN II (as measured
by [O/H]). The field star trend, as defined by both the Milky Way
halo and the dwarf galaxies, suggests a less efficient cycling 
of gas into stars when compared to the metal-poor tail of the
bulge population.  The trend defined
by the highest metallicity bulge stars appears to be a smooth
continuation from the lower metallicities.  The overall
behavior of Na/O is suggestive of 
SN II dominating the nucleosynthesis of Na in the bulge population.

In less massive stars than those that become SN II, Na production
is possible via H-burning on the NeNa-cycle.  This can occur at
temperatures of a few to several times 10$^{7}$K and one byproduct
of this process can be $^{16}$O depletion due to the ON parts of the
CNO-cycles.  In the presence of significant contributions from
NeNa- and CNO-cycles (as an explantion for the large Na abundances
in the bulge) would be an expected anticorrelation of O and Na.
The bottom panel of Figure 6 illustrates just such an effect
where the same Galactic and dwarf galaxy field stars are plotted,
as well as the bulge stars, with the addition of stars from two 
Galactic globular clusters (M13 and M4 as the 4-sided and 3-sided
crosses, respectively).  Here the H-burning signature is clear with
the expected striking Na-O anticorrelations.  No such trend is hinted
at in this (admittedly small) bulge sample, indicating that chemical
evolution in Na and O for the bulge has been driven largely by SN II.

In addition, in the bottom panel of Figure 6 we also show Na and O
results for the metal-rich old open clusters NGC 6791 (Peterson
\& Green 1998) which agrees with the high sodium
found here for the bulge targets. The solid 
curve plotted in this panel represents the Na and O yields from
the SN II models of Woosley \& Weaver (1995) convolved with
a standard, i.e., Salpeter, mass function. The model curve
is suggestive of the trend observed in the field stars, open clusters,
and bulge stars and, within modest differences, points to SN II
enrichment.
It should be mentioned, however, that until more metal-rich
SN II model yields are added to this plot, it may be that
the most metal-rich bulge stars could show some contamination
in Na and O from H-burning in a generation of intermediate mass stars. 
In addition, because the Bulge exhibits an abundance range in Fe, as well as
a large variation in its Na/O values, it may show some similarities
in the behavior of these 3 elements to that of the peculiar globular
cluster $\omega$ Centauri (Norris \& Da Costa 1995; Smith et al. 2000). 

\subsection{Titanium and the Behavior of [Ti/Fe] and [Ti/O] in the
Bulge}

In the majority of stars studied in the Milky Way field halo and disk 
populations titanium behaves as an $\alpha$-element with respect to
iron (e.g. see review by McWilliam 1997).  This statement can be
illustrated in Figure 7 (top panel) with [Ti/Fe] plotted versus
[Fe/H] and the Galactic field stars shown as the small symbols.  The
Galactic results are taken from Gratton \& Sneden (1988); Edvardsson et al.
(1993); McWilliam et al. (1995); Prochaska et al. (2000); Carretta
et al. (2002); Fulbright (2002); Johnson (2002); Bensby et al. (2003);
Reddy et al. (2003).
At metallicities with [Fe/H]$\le$-0.6 to -0.8, [Ti/Fe] is elevated at 
a nearly constant plateau of $\sim$+0.3 dex.  At increasing Fe abundances
there is a rapid, perhaps even abrupt decline in [Ti/Fe] to near-solar
ratios, with a roughly constant [Ti/Fe]$\sim$0.0 over [Fe/H]= -0.5 to
+0.4.

Other populations are also illustrated in the top panel of Figure 7
with the large filled circles being Sculptor (Shetrone et al. 2003;
Geisler et al. 2005) and the large filled squares the LMC (Smith et
al. 2002).  These two dwarf galaxies do not track the Galactic trend,
with both smaller galaxies exhibiting the decline in [Ti/Fe] at
lower values of [Fe/H] when compared to the Milky Way disk and halo
fields.  The bulge stars in Figure 7 are the large filled pentagons
(plotted with error bars) and they define a somewhat different trend
than any of the other populations.  There is a tendancy of the bulge
red giants to continue with elevated [Ti/Fe] values past the downturn
shown by the halo-disk field stars (but the results presented here
are based on a small number of 7 stars).  The behavior of [Ti/Fe]
as a function of [Fe/H] in the Bulge is reminiscent of that of
[O/Fe].  The suggestion would be that the Bulge underwent a more
rapid metallicity enrichment from SN II than the halo, but in the
most metal-rich bulge stars there is a decline in [$\alpha$/Fe]
(found in our sample for both O and Ti).

The bottom panel of Figure 7 shows the same stars as in the top
panel but now plotted as [Ti/O] versus [O/H].  As in Figure 6,
replacing Fe with O in the comparisons removes to some approximation
the role of SN Ia and focusses on the role of SN II.  As in the
comparison of [Na/O] versus [O/H], the various stellar populations
now fall along a similar trend, in this case one of an increase of
[Ti/O] as metallicity, i.e. [O/H], increases.  This behavior was
not expected based on the SN II model yields, which are plotted
in this panel as the solid curve.  This curve is defined by the
Woosley \& Weaver (1995) yields convolved with a Salpeter mass
function at each metallicity.  Given the large number, and varied
natures of the studies plotted here, the scatter in [Ti/O] at a
given [O/H] is not large ($\sim$$\pm$0.25 dex).  The 
lower-metallicity bulge stars studied here fall right in the 
trend defined by the Galactic halo-disk field stars.  The metal-rich
bulge stars continue as an extension of this trend.  The small 
cluster of metal-rich disk stars (with [O/H]$>$0.0) are from
Bensby et al. (2003) and agree well with the metal-rich bulge stars.

Titanium is a product of explosive Si-burning, probably as a result
largely of SN II, with the dominant $^{48}$Ti isotope derived from
the production and subsequent decay of $^{48}$Cr.  The smooth
trend defined by the populations of Sculptor, the LMC, the Galactic
field stars, and the bulge all point to one source and would suggest
SN II.  We offer no explanation for the trend in [Ti/O] but point out 
the significant trend in [Ti/Si] as a function of galactocentric
distance in Galactic globular clusters noted by Lee and Carney (2002).

\section{Conclusions}

$\bf Iron$: 
Our bulge sample, although admittedly small,
spans a significant range across the metallicity distribution as
found from low resolution studies of the bulge. The sampled
metallicity range thus provides the opportunity to infer
characteristics of bulge chemical evolution as defined by
the abundances summarized below.

{\bf Carbon, Nitrogen and the CN cycle}: 
The $^{12}$C-depletion and $^{14}$N-enhancements
are indicative of CN-cycled material. The conclusion of this investigation
into the $^{12}$C and $^{14}$N abundances is that the $^{16}$O abundances
in the studied bulge red giants are not measurably altered from their primordial
values, as the mixing is not nearly extensive enough to have altered
the initial $^{16}$O measurably. Therefore, the derived oxygen abundances
represent good monitors of 
chemical evolution within the bulge population.

$\bf Oxygen$ :
At low metallicity, the [O/Fe] results
agree with halo-like [O/Fe] abundances. The [O/Fe] values then
decline as [Fe/H] increases.
The decline in [O/Fe] in the bulge, however, is not as large as for
the thin and thick disk at the highest metallicities. The most
straightforward explanation of this trend is that the bulge underwent
more rapid metal enrichment than the halo, but that star formation
did continue over timescales that may include the onset of SN Ia.

$\bf Sodium$: 
Near solar metallicity ([O/H] $\sim$ 0.0), the bulge stars
overlap the field-star distributions but the lowest metallicity
bulge star in our sample (IV-003) falls well below
the trend. 
Since the Na/O yields from SN II are metallicity sensitive,
the low value of Na/O for IV-003 may indicate that the initial
enrichment of the bulge to [O/H] $\sim$ -1 
was quite rapid and dominated by very metal-poor massive stars
(with [m/H] less than -2). This would be an example in which
a metallicity-dependant ratio (Na/O) retained the metallicity
signature of its parent SN II.
At the highest metallicities, the sodium abundances are very high,
defining a sharp upward trend with the oxygen abundance. It is interesting
to note that this pattern of high sodium was also found for the metal rich
open cluster NGC 6791 by Peterson \& Green (1998).

$\bf Titanium$: 
The Ti abundances obtained  
define a somewhat different trend than any of the other populations
in the Milky Way. There is a tendancy of the bulge
red giants to retain elevated [Ti/Fe] values past the downturn
shown by the halo-disk field stars. 
In addition, the bulge stars are significantly
more elevated relative to
[Ti/Fe] versus [Fe/H] in dwarf spheroidals.
The behavior of [Ti/Fe]
as a function of [Fe/H] in the bulge is reminiscent of that of
[O/Fe].  The suggestion would be that the bulge underwent a more
rapid metallicity enrichment from SN II than the halo, but in the
most metal-rich bulge stars there is a decline in [$\alpha$/Fe]
(found in our sample for both O and Ti).

\section{Acknowledgements}

We thank Jon Fulbright, Vanessa Hill, Manuela Zoccali and Dante Minniti
for helpful discussions.
Based on observations obtained at the Gemini Observatory,
which is operated by the Association
of Universities for Research in Astronomy, Inc., under a cooperative
agreement with the NSF on behalf of the Gemini partnership: the National
Science Foundation (United States), the Particle Physics and Astronomy
Research Council (United Kingdom), the National Research Council (Canada),
CONICYT (Chile), the Australian Research Council (Australia),
CNPq (Brazil), and CONICRT (Argentina), as program GS-2004A-Q-20.  This
paper uses data obtained with the Phoenix infrared spectrograph,
developed and operated by the National Optical Astronomy Observatory.
This work is also supported in part by the National Science Foundation through
AST03-07534 (VVS), NASA through NAG5-9213 (VVS), and AURA, Inc. through
GF-1006-00 (KC).


\clearpage

\begin{deluxetable}{cccccccc}
\setcounter{table}{0}  
\tablewidth{400pt}  
\tablecaption{Stellar Quantities \& Parameters}
\tablehead{
\colhead{Star} &   
\colhead{A$_{\rm V}$} &
\colhead{V$_{0}$} &
\colhead{J$_{0}$} &
\colhead{K$_{0}$} &
\colhead{T$_{\rm eff}$(K)} &
\colhead{Log g$^{a}$} &
\colhead{$\xi$ (km s$^{-1}$)}
} 
\startdata 
I-322      & 1.48 & 13.01 & 10.75 &  9.93 & 4250 & 1.5 & 2.0 \\ 
IV-003     & 1.30 & 13.70 & 11.75 & 11.03 & 4500 & 1.3 & 1.8 \\
IV-167     & 1.46 & 15.59 & 13.45 & 12.67 & 4375 & 2.5 & 2.2 \\
IV-072     & 1.46 & 14.93 & 12.73 & 11.99 & 4400 & 2.4 & 2.2 \\
IV-329     & 1.35 & 13.81 & 11.54 & 10.73 & 4275 & 1.3 & 1.8 \\
BMB 78     & 1.35 &  --   &  8.52 & 7.38  & 3600 & 0.8 & 2.5 \\
BMB 289    & 1.62 &  --   &  7.29 & 5.98  & 3375 & 0.4 & 3.0 \\
\enddata
\tablecomments{(a): Units of cm s$^{-2}$. 
}
\end{deluxetable}

\clearpage

\begin{deluxetable}{ cccccccc }  
\setcounter{table}{1}  
\tablewidth{430pt}  
\tablecaption{$^{12}$C$^{16}$O Lines and Equivalent Widths$^{a}$ for K-giants}

\tablehead{  $\lambda$(\AA) &   
\multicolumn{1}{c} {$\chi$(eV)} &
\multicolumn{1}{c} {Log gf} &
\multicolumn{1}{c} {I-322} &
\multicolumn{1}{c} {IV-003} &
\multicolumn{1}{c} {IV-072} &
\multicolumn{1}{c} {IV-167} &
\multicolumn{1}{c} {IV-329} 
} 
\startdata 
23350.777 & 0.42  & -5.084 & 579 & 302 & 590 & ---  & ---  \\
23367.117 & 0.01  & -6.338 & 384 & 86  & 394 & 610  & 275  \\
23368.221 & 2.15  & -4.780 & 86  & 54  & 73  & 69   & 75   \\
23372.385 & 0.40  & -5.124 & --- & 260 & 506 & 508  & 405  \\
23373.398 & 1.66  & -4.456 & --- & 113 & 244 & 218  & 189  \\
23383.809 & 0.39  & -5.146 & 527 & 244 & --- & 525  & 394  \\
23385.273 & 1.69  & -4.447 & --- & 69  & 237 & 255  & 194  \\
23386.213 & 0.00  & -6.438 & --- & 112 & 332 & 302  & 242  \\
23388.523 & 2.20  & -4.772 & 96  & --- & --- & ---  & ---  \\
23395.646 & 0.38  & -5.167 & 547 & 255 & 517 & 448  & 382  \\
23397.609 & 1.73  & -4.439 & 350 & 62  & --- & 180 & 158  \\
23407.893 & 0.36  & -5.190 & 535 & 234 & 507 & 620 & 387  \\
23410.404 & 1.76  & -4.431 & 247 & 50  & 303 & 380 & 161  \\
23420.557 & 0.35  & -5.214 & 510 & --- & --- & 462 & 440  \\
23423.668 & 1.80  & -4.421 & 252 & --- & 278 & 285 & 196  \\
23425.664 & 0.00  & -6.745 & 307 & --- & 222 & 225 & 215  \\
23430.629 & 2.29  & -4.757 & 79  & --- & --- & --- & ---  \\
23433.635 & 0.35  & -5.240 & 520 & --- & 460 & --- & 411  \\
23437.393 & 1.84  & -4.414 & --- & --- & 268 & 268 & ---  \\

\enddata
\tablecomments{(a): Equivalent widths in m\AA.  
}
\end{deluxetable}

\clearpage

\begin{deluxetable}{ ccc }
\setcounter{table}{2}
\tablewidth{430pt}
\tablecaption{Lines used as Abundance Indicators via Spectrum Synthesis}

\tablehead{ $\lambda$(\AA) &
\multicolumn{1}{c} {$\chi$(eV)} &
\multicolumn{1}{c} {log gf} 
}
\startdata
   Fe I    &      &        \\
15531.742 & 5.64 & -0.564 \\
15531.750 & 6.32 & -0.900 \\
15534.239 & 5.64 & -0.402 \\
15537.572 & 5.79 & -0.799 \\
15537.690 & 6.32 & -0.660 \\
  Na I    &      &        \\
23378.945 & 3.75 & -0.420 \\
23379.139 & 3.75 & +0.517 \\
  Ti I    &      &        \\
15543.720 & 1.88 & -1.481 \\
$^{12}$C$^{14}$N & &      \\
15530.987 & 0.89 & -1.519 \\
15544.501 & 1.15 & -1.146 \\
15552.747 & 0.90 & -1.680 \\
15553.659 & 1.08 & -1.285 \\
15563.376 & 1.15 & -1.141 \\
$^{16}$OH &      &        \\
15535.489 & 0.51 & -5.233 \\
15560.271 & 0.30 & -5.307 \\
15568.807 & 0.30 & -5.270 \\
15572.111 & 0.30 & -5.183 \\
\enddata
\end{deluxetable}

\clearpage

\begin{deluxetable}{ ccccccc }  
\setcounter{table}{3}  
\tablewidth{420pt}  
\tablecaption{Abundances$^{a}$}

\tablehead{  Star &   
\multicolumn{1}{c} {A($^{12}$C)} &
\multicolumn{1}{c} {A($^{14}$N)} &
\multicolumn{1}{c} {A($^{16}$0)} &
\multicolumn{1}{c} {A(Na)} &
\multicolumn{1}{c} {A(Ti)} &
\multicolumn{1}{c} {A(Fe)} 
} 
\startdata 
I-322    & 7.90 & 8.20 &  8.60   & 6.13 &  5.15  & 7.21 \\
IV-003   & 7.00 & 7.25 &  8.05   & 4.23 &  4.14  & 6.40 \\
IV-072   & 8.70 & 8.62 &  9.20   & 7.35 &  5.44  & 7.69 \\
IV-167   & 8.44 & 8.40 &  9.10   & 7.30 &  5.27  & 7.82 \\
IV-329   & 7.30 & 7.60 &  8.35   & 5.30 &  4.64  & 6.93 \\
BMB78    & 8.30 & 8.50 &  9.00   &  --  &  5.05  & 7.42 \\
BMB289   & 8.20 & 8.55 &  8.75   &  --  &  4.94  & 7.40 \\
Arcturus & 7.92 & 7.60 &  8.49   & 5.85 &  4.83  & 6.96 \\ 
\enddata
\tablecomments{(a): A(X)= log[n(X)/n(H)] + 12. The solar iron abundance
is A(Fe)$_{\odot}$= 7.45.
}
\end{deluxetable}

\begin{deluxetable}{ cccccc }
\setcounter{table}{4}
\tablewidth{420pt}
\tablecaption{Abundances Errors due to Stellar Parameter Uncertainties}

\tablehead{ Element &
\multicolumn{1}{c} {$\delta$T=+75K} &
\multicolumn{1}{c} {$\delta$log g=+0.25} &
\multicolumn{1}{c} {$\delta \xi$=+0.3km s$^{-1}$} &
\multicolumn{1}{c} {$\Delta ^{a}$} & 
\multicolumn{1}{c} {$\delta$T + $\delta$log g $^{b}$}
}
\startdata
$^{12}$C & +0.05  & +0.12 &  -0.08  &  $\pm$0.15 & -0.12 \\
$^{14}$N & +0.08  & +0.10 &  -0.06  &  $\pm$0.14 & -0.17 \\
$^{16}$O & +0.12  & +0.12 &  -0.03  &  $\pm$0.17 & -0.19 \\
Na       & +0.10  & +0.01 &  -0.02  &  $\pm$0.10 & -0.13 \\
Ti       & +0.12  & +0.02 &  -0.09  &  $\pm$0.15 & -0.10 \\
Fe       & +0.01  & +0.05 &  -0.11  &  $\pm$0.12 & -0.01 \\
\enddata
\tablecomments{(a): [($\delta T)^{2}$ + ($\delta$log g)$^{2}$ + ($\delta \xi)^{2}$]$^{1/2}$ 
}
\tablecomments{(b): Simultaneous change in $\delta$ T=-75K and $\delta$log g=-0.15. This changed $\xi$
from 2.2 to 2.1 km/s.
}
\end{deluxetable}

\clearpage

\begin{figure}\plotone{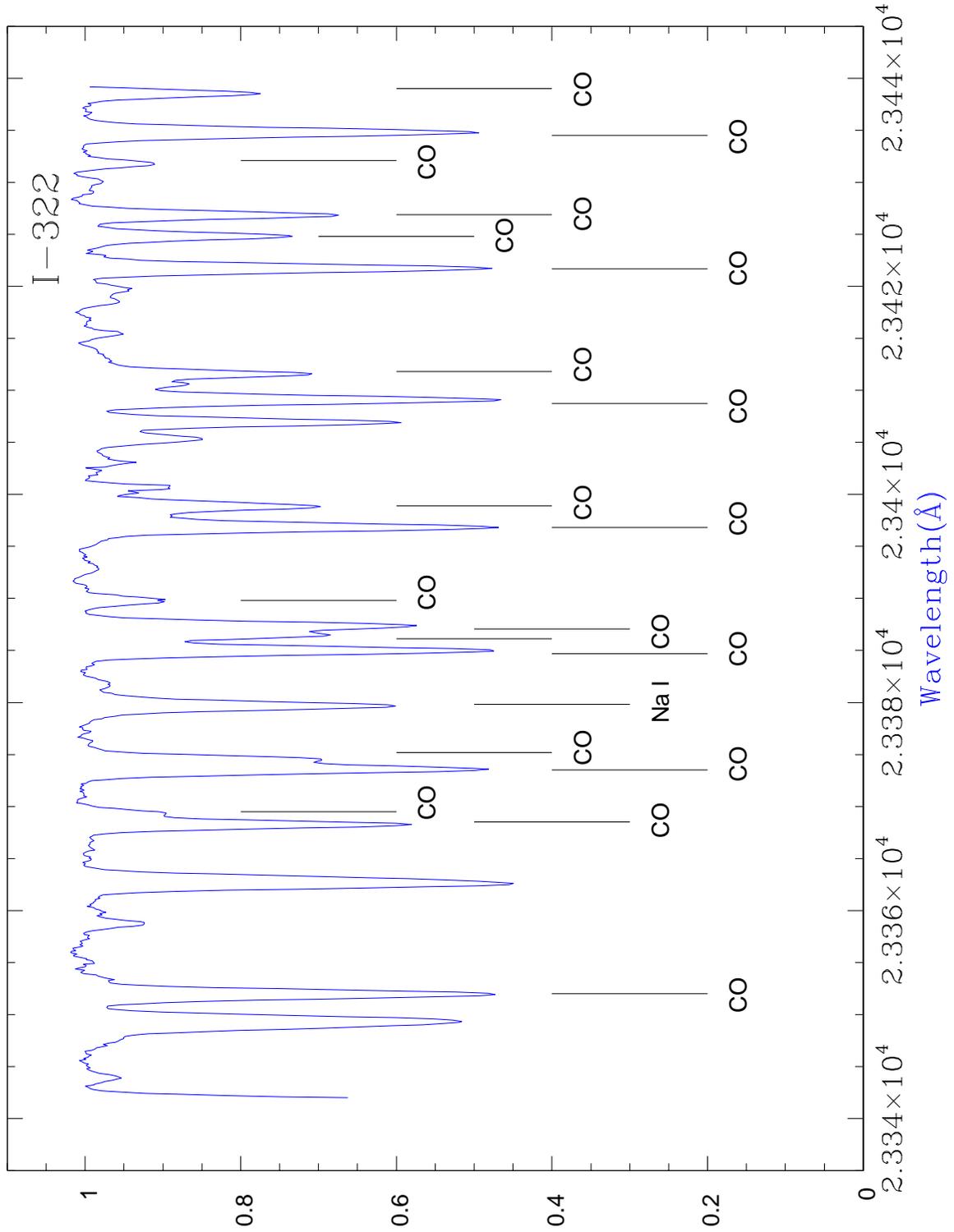}
\figcaption[f1.ps]{Observed spectrum in the K-band region of target star
I-322. The selected CO molecular features (Table 2) are labelled, as well
as the studied Na I line at 233379.137\AA.
\label{fig1}}\end{figure}

\clearpage

\begin{figure}
\epsscale{.8}\plotone{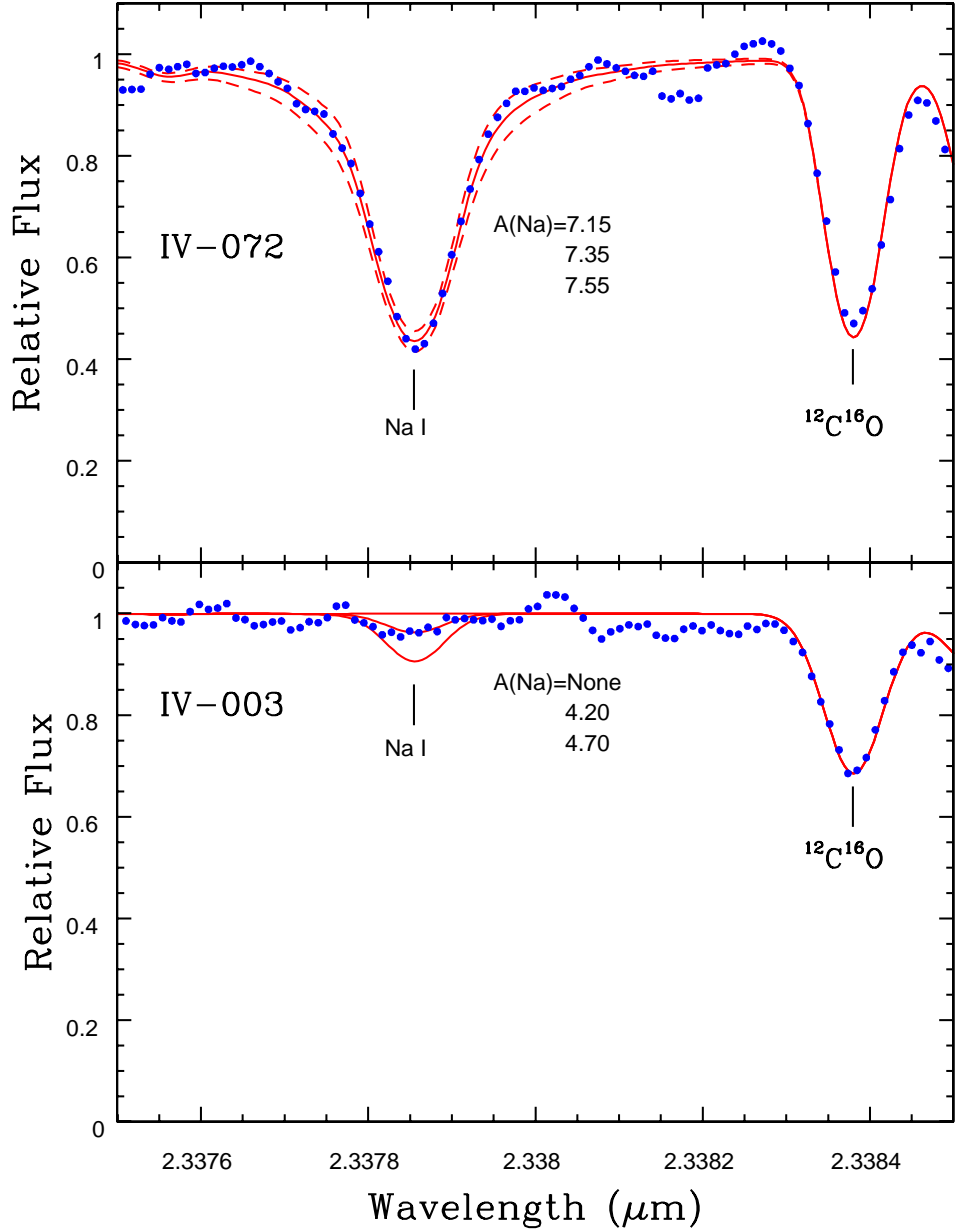} 
\figcaption[f2.eps]{Observed and synthetic spectra for the two bulge
giants in our sample which illustrate the large spread in Na abundances.
We show Na I fits for one bulge target with a large Na abundance and another with
sodium abundances much lower (close to an upper limit detection). 
The latter falls well below 
the trend of [Na/O] vs [O/H] defined by field stars in the Milky Way. 
\label{fig2}}\end{figure}

\clearpage

\begin{figure}\plotone{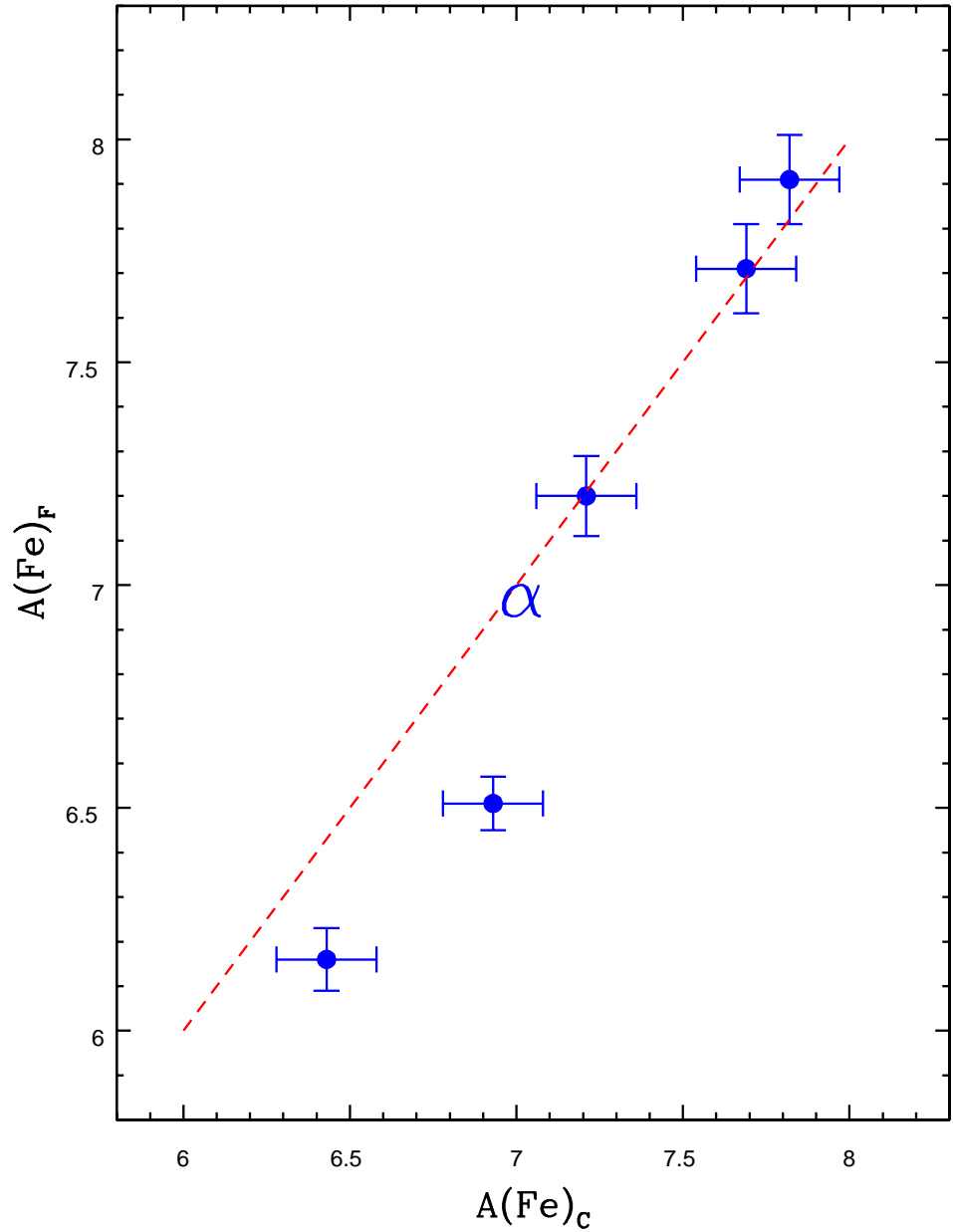}
\figcaption[f3.eps]{A comparison between the Fe abundances obtained
for the 5 bulge K-giants in common with Fulbright et al. (2006) sample.
The dashed line represents perfect agreement. The differences between
the two studies are close to the expected uncertainties in each one,
although there may be a metallicity dependent difference. We also
show for compariosn the derived abundances in the two studies for the reference 
star Alpha Boo (represented by the $\alpha$ symbol; Table 4).
\label{fig3}}\end{figure}

\begin{figure}\plotone{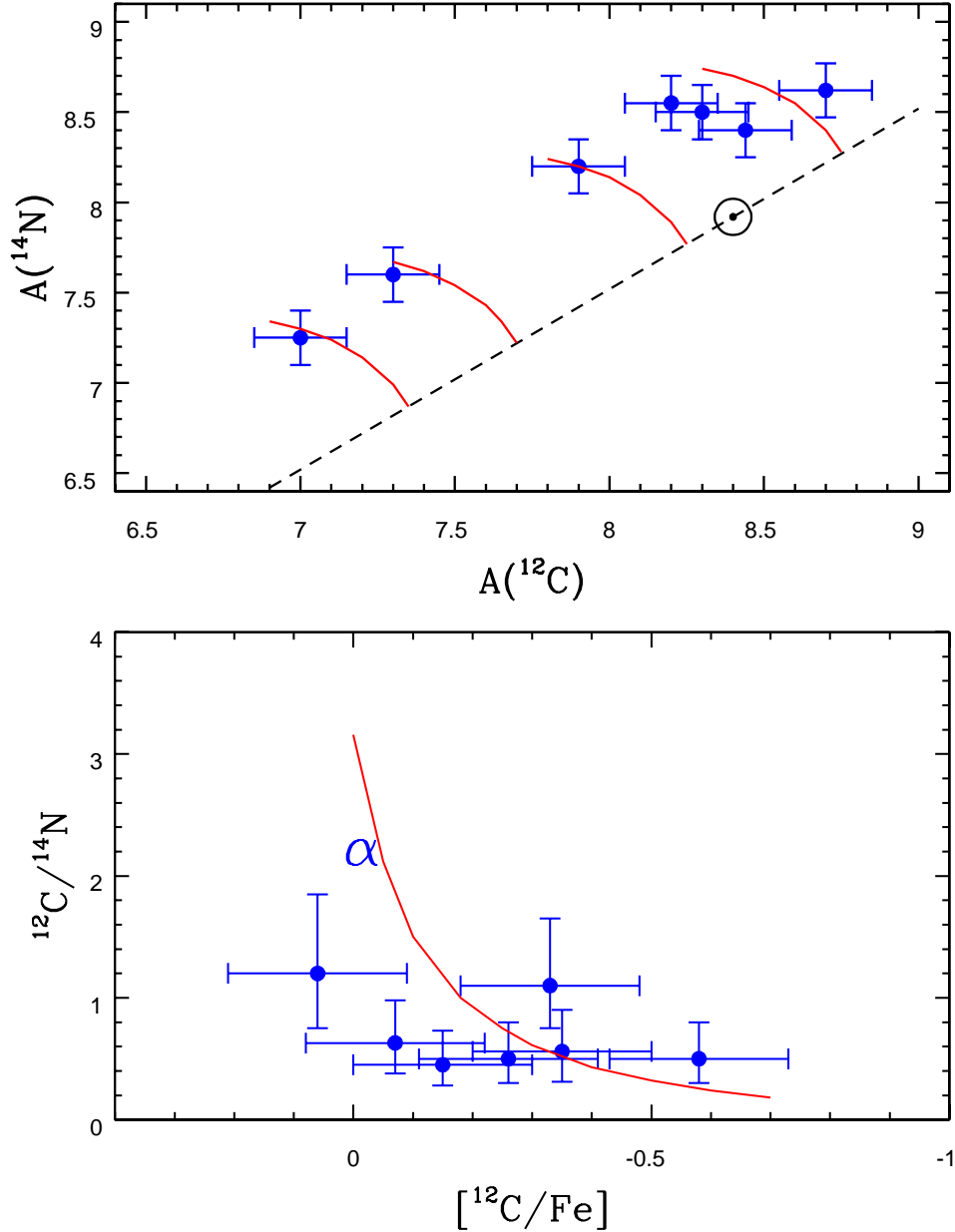}
\figcaption[f4.eps]{The top panel shows the nitrogen (A($^{14}$N)) versus 
carbon (A($^{12}$C)) abundances for the sample stars.
The dashed line delineates the scaled solar line: all
studied bulge red giants fall in the $^{14}$N-rich zone. 
The solid curves show lines of constant
$^{12}$C+$^{14}$N, or the so-called 'CN-mixing lines'. 
The bottom panel shows a CN-mixing curve normalized to the Fe abundance,
so that all of the bulge red giants can be plotted with respect to a single
curve. A solar $^{12}$C/$^{14}$N=3.16
is taken as the initial value and declines 
as [$^{12}$C/Fe] declines. Alpha Boo is also plotted (represented by the $\alpha$ symbol)
with its $^{12}$C, $^{14}$N, and Fe abundances derived from the same lines as used for
the bulge giants. 
\label{fig4}}\end{figure}

\clearpage

\begin{figure}\plotone{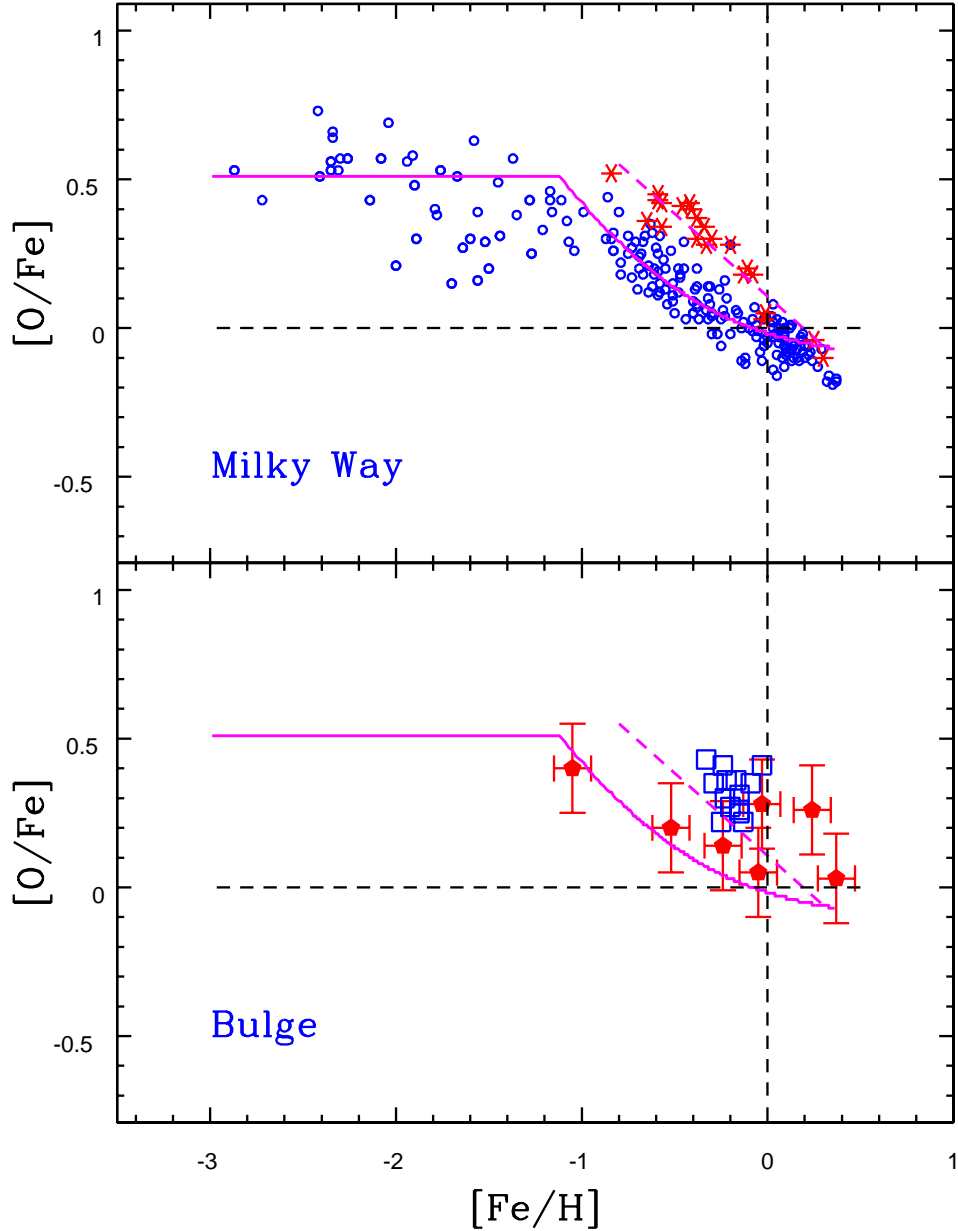}
\figcaption[f5.eps]{The behavior of [O/Fe] versus [Fe/H] for the Galactic disk 
and halo populations, along with the bulge. The top panel shows [O/Fe] from
a number of studies (identified in the text) for halo and disk stars.
The asterisks are specifically thick disk stars from Bensby et al. (2004)
which are displayed in [O/Fe] relative to thin disk stars. The solid
curve is a chemical evolution model for O and Fe yields from SN II and SN Ia,
while the dashed line is a linear fit to the thick disk stars. The bottom
panel illustrates [O/Fe] for bulge red giants. This study's stars are
shown as the large filled pentagons (with error bars) while the M-giants
from Rich \& Origlia (2005) are plotted as open squares. The galactic halo and
disk curves from the top panel are also plotted as comparisons.
\label{fig5}}\end{figure}

\clearpage

\begin{figure}\plotone{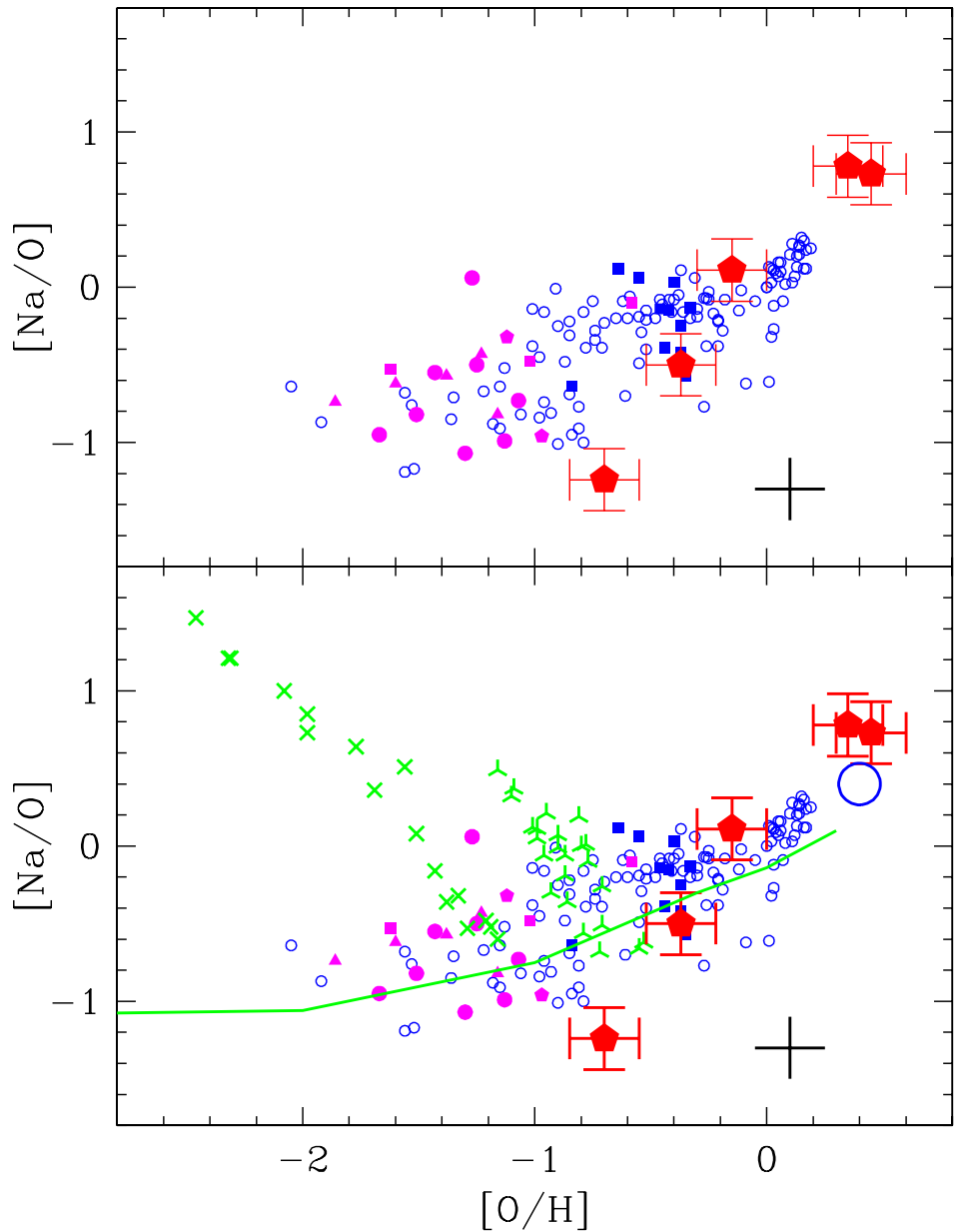}
\figcaption[f6.eps]{The behavior of Na and O in various stellar
populations. Open circles are Galactic halo and disk results from
many studies (noted in the text) while the small filled
symbols represent Sculptor (circles), Carina (triangles), Fornax \& Leo
(pentagons), and the LMC (squares). This variety of stellar populations
loosely define an increasing [Na/O] with [O/H]. An increasing
Na/O yield with increasing oxygen is the qualitative prediction for
SN II; model yields from Woosley \& Weaver (1995) are added (the solid
curve) in the bottom panel. Included in the bottom panel are abundances
from the globular cluster M15 (4-point crosses) and M4 (3-point crosses);
the Na-O anti-correlations are the signature of H-burning. The metal-rich
open cluster NGC 6791 is also shown as the large open circle
(from Peterson \& Green 1998) and this metal rich point agrees with our
most metal-rich bulge stars. 
\label{fig6}}\end{figure}

\begin{figure}\plotone{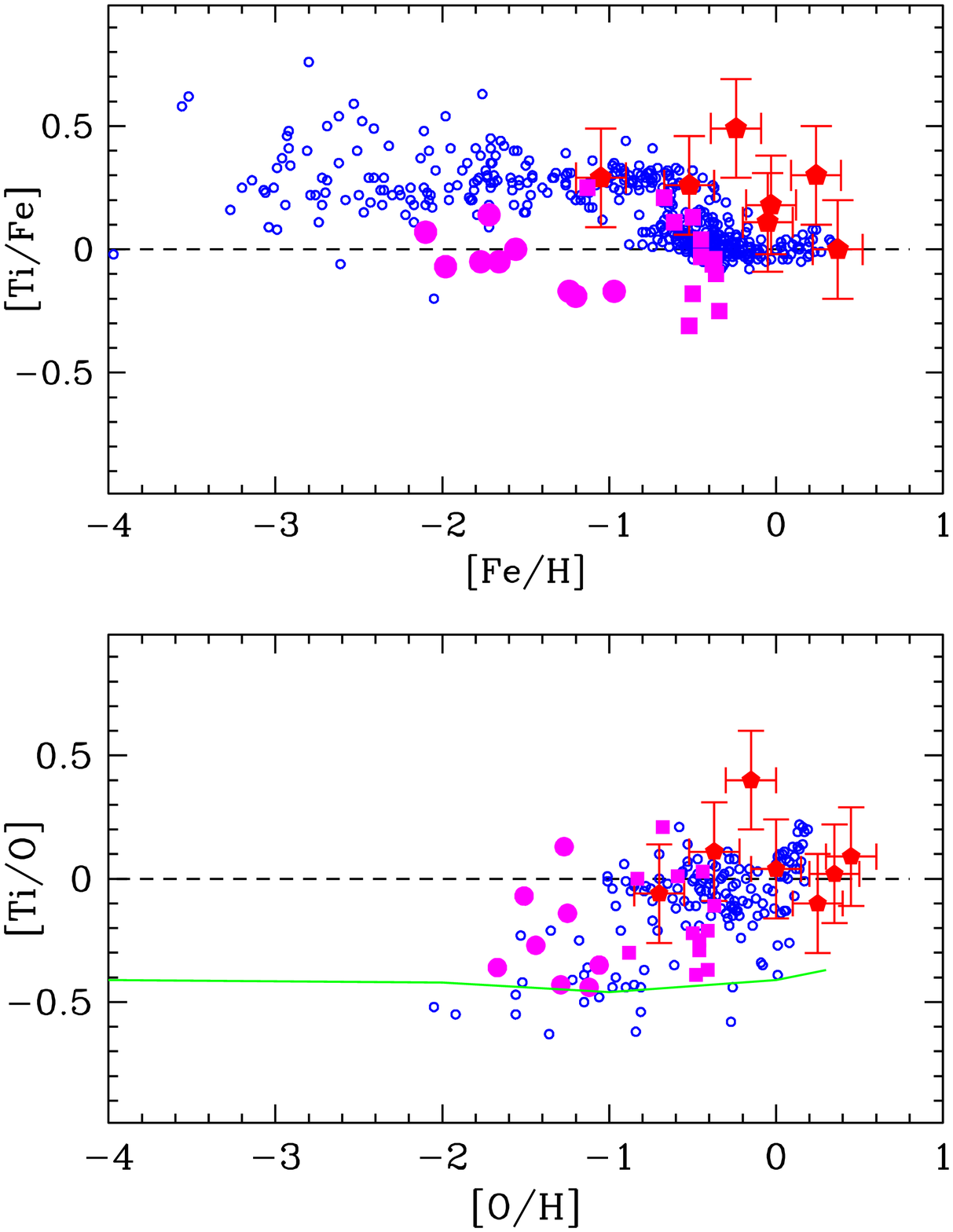}
\figcaption[f7.eps]{The top panel shows [Ti/Fe] versus
[Fe/H] for different populations. The Galactic field stars shown as the small open symbols
and are taken from a number of studies identified in the text.
Other populations are also illustrated: large filled circles Sculptor (Shetrone et al. 2003;
Geisler et al. 2005); large filled squares the LMC (Smith et al. 2002).
The bulge stars are shown as the large filled pentagons (with error bars). As with
[O/Fe], there is a tendency for the bulge values of [Ti/Fe] to remain elevated
somewhat relative to the disk.
The bottom panel shows the same stars as in the top
panel but now plotted as [Ti/O] versus [O/H]. 
A general increase of [Ti/O] as [O/H] increases is seen as defined by all of
the stellar populations. This increase is not predicted by the Woosley \& Weaver
(1995) yields that are shown as the solid curve in the bottom panel.
\label{fig7}}\end{figure}


\begin{thebibliography}{}

\bibitem[]{825} Alonso, A, Arribas, S. \& Martinez-Roger, C. 1999, A\&A Supp. 140, 261
\bibitem[]{826} Asplund, M., Grevesse, N., \& Sauval, A. J. 2005 In Cosmic
Abundances as Records of Stellar Evolution and Nucleosynthesis, ed. F. N. Bash, \&
T. G. Barnes p. 25
\bibitem[]{829} Baade, W. 1944, ApJ, 100, 137
\bibitem[]{830} Bensby, T., Feltzing, S., \& Lundstrom, I. 2003, A\&A, 410, 527
\bibitem[]{831} Bensby, T., Feltzing, S., \& Lundstrom, I. 2004, A\&A, 415, 155
\bibitem[]{832} Bessel, M.S., Castelli, F., \& Plez, B. 1998, A\&A, 333, 231
\bibitem[]{824} Cayrel, R. 1988, in IAU Symp. 132, The Impact of Very High S/N Spectroscopy on Stellar Physics, ed. R. Cayrel, G. de Strobel \& M. Spite (Dordrecht: Kluwer), 354 
\bibitem[]{833} Carpenter, J. M. 2001, ApJ, 121, 2851 
\bibitem[]{834} Carretta, E., Gratton, R., Cowan, J.G., Beers, T.C. ,\& Cristlieb, N. 2002, AJ, 124, 481
\bibitem[]{835} Clayton, D. 2003, Handbook of Isotopes in the Cosmos (Cambridge University Press: Cambridge)
\bibitem[]{836} Costes, M., Naulin, C., \& Dorthe, G. 1990, A\&A, 232, 270
\bibitem[]{837} Cunha, K., Smith, V., \& Lambert, D. L. 1998, ApJ, 493, 195
\bibitem[]{838} Cunha, K., Smith, V. V.; Suntzeff, N. B.; Norris, J. E.; Da Costa, G. S.; Plez, B.	2002, AJ, 124, 379
\bibitem[]{839} Edvardsson, B., Andersen, J., Gustafsson, B., Lambert, D. L., Nissen, P. E., \& Tomkin, J. 1993, A\&A, 275, 101
\bibitem[]{840} Fulbright, J. P. 2000, AJ, 120, 1841
\bibitem[]{841} Fulbright, J. P. 2002, AJ, 123, 404
\bibitem[]{842} Fulbright, J. P. \& Johnson, J.A. 2003, ApJ, 595, 1154
\bibitem[]{843} Fulbright, J., McWilliam, A., \& Rich, M. R. 2006 ApJ, 636, 821
\bibitem[]{844} Girardi, L. Bressan, A. Bertelli, G., \& Chiosi, C. 2000, A\&AS, 141, 371
\bibitem[]{845} Geisler, D., Smith, V. V., Wallerstein, G., Gonzalez, G. \& Charbonnel, C. 2005, AJ, 129,1428 
\bibitem[]{846} Goldman, A., Shoenfeld, W. G., Goorvitch, D., Chackerian, C. Jr., Dothe, H., Melen, F., Abrams, M.C., \& Selby, J. E. A. 1998, J. Quant. Spectrosc. Radiat. Transfer 59, 453 
\bibitem[]{847} Goorvitch, D. 1994, ApJS, 95, 535
\bibitem[]{848} Gratton, R. G., \& Sneden, C. 1988, A\&A, 204, 193
\bibitem[]{849} Gratton, R. G., Bonifacio, P., Bragaglia, A., Carretta, E., Castellani, V., Centurion, M., Chieffi, A., Claudi, R., Clementini, G., D'Antona, F., and 9 coauthors 2001, A\&A, 369, 87
\bibitem[]{850} Gustafsson, B., Bell, R. A., Eriksson, K., \&  Nordlund, A. 1975, A\&A, 42, 407 
\bibitem[]{851} Hinkle, K. H., Wallace, L., \& Livingston, W. 1995, Infrared Atlas of
the Arcturus Spectrum, 0.9--5.3Z$\mu$m (San Francisco:ASP)
\bibitem[]{853} Hinkle et al. 2003, Proc. SPIE, 4834, 353
\bibitem[]{854} Houdashelt, M. L., Bell, R. A., \& Sweigart, A. V. 2000, AJ, 119, 1448
\bibitem[]{855} Huber, K-P, \& Herzberg, G.  1979, Constants of Diatomic Molecules (New York: Van Nostrand)
\bibitem[]{856} Johnson, J. A. 2002, ApJS, 139, 219
\bibitem[]{857} Kurucz, R. \& Bell, R. 1995, Atomic spectral line database from CD-ROM 23
\bibitem[]{858} Lee, J., \& Carney, B. W. 2002, AJ, 124, 1511 
\bibitem[]{859} McWilliam, A. \& Rich, R. M.  1994, ApJS, 91, 749
\bibitem[]{860} McWilliam, A., Preston, G. W., Sneden, C., Searle, L. 1995, AJ, 109, 2757
\bibitem[]{861} Melendez, J. \& Barbuy, B. 1999, ApJS, 124, 527
\bibitem[]{862} Melendez, J. \& Barbuy, B. 2002, ApJ, 575, 474
\bibitem[]{863} Melendez, J. \& Barbuy, B., \& Spite, F. 2001, ApJ, 556, 858
\bibitem[]{864} Nissen, P.E., \& Schuster W.J 1997, A\&A, 326, 751
\bibitem[]{865} Nissen, P.E., Primas, F., Asplund, M., \& Lambert, D. L. 2002, A\&A, 390, 235
\bibitem[]{820} Norris, J. E., \& Da Costa, G. S. 1995, ApJ, 447, 680 
\bibitem[]{867} Peterson, R. C., \& Green, E. M. 1998, ApJ, 502, L39 
\bibitem[]{868} Prochaska, J. X., Naumov, S. O., Carney, B. W., McWilliam, A., Wolfe, A. M. 2000, AJ, 120, 2513
\bibitem[]{869} Plez, B., Brett, J.M. \& Nordlund, A. 1992 A\&A, 256, 551
\bibitem[]{870} Ramirez, I. \& Melendezi, J. 2005, ApJ, 626, 465
\bibitem[]{871} Reddy, B. E., Tomkin, J., Lambert, D. L., \& Allende Prieto, C. 2003, MNRAS, 340, 304
\bibitem[]{872} Rich R. M. \& Whitford, A. E. 1983, ApJ, 274, 723
\bibitem[]{873} Rich, R. M. 1988, AJ, 95, 828
\bibitem[]{874} Rich, R. M. 1990, ApJ, 362, 604
\bibitem[]{866} Rich, R. M. \& Origlia, L. 2005, ApJ, 634, 1293 
\bibitem[]{875} Sadler, E. M., Rich, R. M.,\& Terndrup, D. M. 1996, AJ, 112, 171
\bibitem[]{876} Shetrone, M., Venn, K. A., Tolstoy, E., Primas, F., Hill, V., Kaufer, A. 2003, AJ, 125, 684
\bibitem[]{877} Sneden, C. 1973, ApJ, 184, 839
\bibitem[]{878} Smith, V. V. \& Lambert, D. L. 1985, ApJ, 303, 226 
\bibitem[]{879} Smith, V. V. \& Lambert, D. L. 1986, ApJ, 311, 843 
\bibitem[]{880} Smith, V. V. \& Lambert, D. L. 1990, ApJS, 72, 387
\bibitem[]{881} Smith, V. V., Suntzeff, N. B., Cunha, K., Gallino, R., Busso, M., Lambert, D. L., Straniero, O 2000, AJ, 119, 1239
\bibitem[]{882} Smith, V. V., Cunha, K., \& King, J. R. 2001, AJ, 122, 370
\bibitem[]{883} Smith, V. V., Hinkle, K. H., Cunha, K., Plez, B., Lambert, D. L., Pilachowski, C. A., Barbuy, B., Melendez, J., Balachandran, S., Bessel, M. S., Geisler, D. P., Hesser, J. E., \& Winge, C., 2002, AJ, 124, 3241
\bibitem[]{884} Smith, V. V.,  Cunha, K., Ivans, I., Lattanzio, J. C.; Campbell, S., Hinkle, K. H. 2005, ApJ, 633, 392
\bibitem[]{885} Stanek, K. Z. 1996, ApJ, 460, L37
\bibitem[]{886} Terndrup, D. M., Sadler, E. M., \& Rich, R. M. 1995, AJ, 110, 1774
\bibitem[]{887} Udalski, A.; Szymanski, M.; Kaluzny, J.; Kubiak, M.; Mateo, M. 1993, Acta Astron., 43, 69
\bibitem[]{888} Udalski, A.; Szymanski, M.; Stanek, K. Z.; Kaluzny, J.; Kubiak, M.; Mateo, M.; Krzeminski, W.; Paczynski, B.; Venkat, R. 1994, Acta Astron., 44, 165
\bibitem[]{889} Ventura, P., \& D'Antona, F. 2005, ApJ, 635, L149
\bibitem[]{890} Wallace, L.; Livingston, W.; Hinkle, K.; Bernath, P. 1996, ApJS, 106, 165
\bibitem[]{891} Woosley, S. E., \& Weaver, T. A. 1995, ApJS, 101, 181


\end{thebibliography}
\end{document}